# Evidence for hidden order and its potential correlation with electron–hole asymmetry in iridate superlattices

***Authors:*** *Zhihan Qiao[1,2,†], Han Zhang[1,3,†], Hangtian Wang[1,2], Deyang Li[1,2], Haiyang Zhang[1,2], Enyang Men[1,2], Junlei Jiang[1,2], Jingxin Chen[1,2], Yizhou Liu[1], Sang-Youn Park[4], Xiangnan Xie[5], MingLiang Tian[1], Haifeng Du[1], Ning Hao[1], Lin Hao[1,*]*

**Affiliations:**

[1]Anhui Provincial Laboratory of Low-Energy Quantum Materials and Devices, High Magnetic Field Laboratory, HFIPS, Chinese Academy of Sciences, Hefei, Anhui 230031, China

[2]Science Island Branch of Graduate School, University of Science and Technology of China, Hefei 230026, China

[3]Wang Zheng School of Microelectronics, Changzhou University, Changzhou, Jiangsu 213001, China

[4]Pohang Accelerator Laboratory, POSTECH, Pohang, Republic of Korea

[5]College of Science, National University of Defense Technology, Changsha 410073, China

**Corresponding Author**

*Lin Hao (Email: haolin@hmfl.ac.cn).

Abstract:

Hidden order refers to an order that shapes the properties of quantum materials yet remains difficult to identify. Here we report signatures of a hidden order in doped iridate superlattices through an unexpected electron-hole asymmetry in the coupling between the insulating state and the antiferromagnetic (AFM) state. While carrier doping suppresses both states, they remain coupled in the electron-doped regime but decouple in the hole-doped regime. Structural analysis rules out octahedral rotation and tilting as the dominant origin of this asymmetry. Instead, magnetic characterization reveals evidences of a hidden order that only appears in the hole-doped regime. This hidden order emerges above $T_N$ and is closely associated with the AFM state, coinciding with an exceptionally strong magnetic scattering intensity and an anomalously robust magnetic coherence. Our results highlight that hole-doped iridates can be a promising platform for exploring hidden and unconventional ordering phenomena in spin-orbit-entangled quantum materials.

## Introduction

Hidden order represents one of the most intriguing phenomena in condensed matter physics, where a phase transition occurs without a clearly determined order parameter[1,2]. Such states have been reported in a variety of quantum materials, where strong electron correlation and/or strong spin-orbit coupling are usually present, yet the microscopic origin often remains elusive[3-7]. When the multiple interactions are delicately balanced, higher-rank order may emerge whose order parameter is either too weak to be observed or only couples weakly to conventional probes[8]. As a result, the order remains *hidden*, even though it can shape the electronic and magnetic properties of the system in an unconventional manner. Understanding hidden order is therefore crucial for revealing emergent phenomena beyond the explicit symmetry-breaking paradigm.

Strongly spin-orbit-coupled iridates provide a particularly attractive platform for exploring hidden order relevant phenomena[2]. In these materials, strong spin-orbit coupling entangles spin and orbital degrees of freedom and stabilizes a $J_{\mathrm{eff}} = 1/2$ state, in which pseudospin, charge and lattice are intimately intertwined[9-13]. Although the ground-state manifold is simplified to the $J_{\mathrm{eff}} = 1/2$ doublet[14,15], various complex hidden phases have been reported[16-20]. This includes a recently identified quadrupole order in the pristine $Sr_2IrO_4$[19]. Another hidden order has been proposed or observed in pristine and carrier-doped $Sr_2IrO_4$[16]. The hidden order was even suggested to be different if the carrier type of dopant varies[17,21]. Despite these advances, the role of hidden order in shaping the magnetic and electronic responses of iridates remains largely unexplored. In particular, how hidden order evolves with a systematic carrier doping also remains an open question.

Here we report a pronounced electron-hole asymmetry in iridate superlattices (SLs), manifested as an unequal coupling between the insulating strength and the antiferromagnetic (AFM) ordering temperature $T_N$ in the electron- and hole-doped regimes. Systematic structural characterization rules out octahedral rotation and tilting as the dominant origin of this asymmetry. Instead, resonant elastic X-ray scattering (REXS) measurements reveal that magnetic coherence is anomalously robust upon hole doping, while it rapidly degrades in the electron-doped regime. Moreover, we identify clear signatures of a hidden order that emerges above $T_N$ exclusively in the hole-doped regime. This hidden order is closely associated with the AFM state and promotes the electron-hole asymmetry, potentially revealing a previously unexplored role of hidden order in governing the electron transport of spin-orbit-coupled Mott systems.

**Results and Discussion**

**Electron-hole asymmetric coupling between insulating strength and $T_N$**

The inset of Fig. 1(a) presents the crystal structure of the SLs used in this study. Each SL consists of identical $IrO_2$ square lattices in neighboring iridate layers without any translational shift. It has been established that this artificial structure effectively captures the quasi-2D $J_{eff} = 1/2$ physics[22-25]. The advantage of this design is that the spin-canting induced net magnetization is always parallel to the neighbors, leading to a spontaneous magnetization in the AFM state[23,25,26]. Due to this inherent coupling, one can probe the AFM order via magnetization measurement[15].

To implement electron and hole doping, we have prepared a series of $(La_xSr_{1-x}IrO_3)1/(SrTiO_3)1$ and $(SrIr_{1-y}Rh_yO_3)1/(SrTiO_3)1$ SLs, respectively (see Supplementary Notes 1 and 2). The electron doping ratio equals to the chemical substitution level of La (x)[27], while the hole doping ratio was calculated follow an established relation between chemical substitution level of Rh (y) and the effective hole doping ratio in Rh-doped $Sr_2IrO_4$[28]. To simplify the discussion, we hereafter adopt the carrier doping ratio rather than the chemical composition substitution level. The maximum electron-doping and hole-doping ratios exploited in this work are -8% and 14%, respectively. The X-ray diffraction (XRD) patterns were measured using a Panalytical X'Pert MRD diffractometer. Electric transport properties were investigated with a physical properties measurement system (Quantum Design). Magnetic properties were measured with a superconducting quantum interference device (SQUID) magnetometer (Quantum Design). Synchrotron XRD (SXRD) and REXS experiments were performed at BL3A beamline of the Pohang accelerate laboratory (PAL) and the 3A beamline of the Photon Factory (Proposal No. 2023G509).

Figure 1(a) displays the temperature-dependent electrical transport properties of the electron- and hole-doped SLs. The resistivity of the pristine SL increases by approximately three orders of magnitude upon cooling to the base temperature, consistent with its spin-orbit Mott insulating nature[25,29]. The resistivity increase is substantially reduced by both electron doping and hole doping. Furthermore, a resistivity anomaly, attributed to the strong spin-charge coupling, was observed in the pristine SL, in agreement with previous reports[26]. This anomaly gradually diminishes upon doping, indicating a concurrent modulation of the magnetic order. To characterize the AFM ordering temperature $T_N$, we performed magnetization measurements. Similar to insulating strength, $T_N$ was also systematically reduced with doping (Fig. 1(b)).

To quantify the insulating strength, we extracted the relative increase of resistivity from room temperature to a specific low temperature by $\Delta = \ln[\frac{\rho(10K)}{\rho(300K)}]$, which effectively characterizes the charge gap[30]. Figure 1(c) presents the doping-dependent phase diagram, highlighting a striking electron-hole asymmetry in two aspects. Firstly, while $T_N$ and $\Delta$ are reduced in both the electron-doped and hole-doped SLs, both parameters are suppressed much more rapidly in the former as compared to the latter. Secondly, we observed different couplings between $T_N$ and $\Delta$ in different regimes. Explicitly, $\Delta$ almost follows $T_N$ in the electron-doped regime, whereas their variations clearly deviate from each other in the hole-doped regime.

Electron-hole asymmetry is well-established in cuprates, and a similar asymmetry is expected in iridates given the close correspondence between the $J_{\mathrm{eff}} = 1/2$ and $S = 1/2$ Mott physics[22-25]. However, in contrast to the well-known electron-hole asymmetry of the superconducting phase diagram in cuprates, here the electron-hole asymmetry emerges within the AFM Mott state itself. More importantly, this asymmetry is not simply manifested as an unequal suppression of either $T_N$ or $\Delta$ on the two sides of doping, but rather emerges as a distinct coupling between $T_N$ and $\Delta$ in the electron- and hole-doped regimes. By reanalyzing available data for electron- and hole-doped cuprates through *A*-site substitution[31-34], we identify a qualitatively similar electron-hole asymmetric relationship between $T_N$ and $\Delta$, but with an opposite carrier polarity, which may relate to the opposite sign of next-nearest-neighbor hopping in the two systems[35,36].

**Asymmetry from the structural point of view**

Electron doping on the *B*-site of iridates remains experimentally challenging and often is under debate[37,38]. Hole doping on the *A*-site with alkali metals[39] is incompatible with high-pressure growth of films. Consequently, electron doping is realized by substituting the *A*-site ions, whereas hole doping is achieved through partial substitution of the *B*-site ions in this work. These chemically distinct routes could, in principle, introduce different structural effects and thereby contribute to an electron-hole asymmetry. To examine this possibility, we investigated averaged lattice distortion using SXRD by probing half-order Bragg reflections[40,41]. In particular, the (0.5 1.5 1.5) and (0.5 0.5 1.5) reflections are signatures of octahedral rotation and octahedral tilting, respectively.

As shown in Fig. 2(a), a strong (0.5 1.5 1.5) Bragg reflection was observed in all the SLs, indicating that the $IrO_6$ octahedral rotation remains robust upon chemical substitution. The normalized intensity of this peak is nearly unchanged across the series (Fig. 2(b)). Because the octahedral rotation is closely related to the in-plane Ir-O-Ir bond angle and is governed by the in-plane lattice structure, this robustness is consistent with the identical in-plane lattice across the SL series imposed by coherent strain from the $SrTiO_3$ substrate (Supplementary Note 3). In contrast, the (0.5 0.5 1.5) reflection is weak in the pristine and hole-doped SLs but becomes significantly enhanced in the electron-doped regime (Fig. 2(c)). As summarized in Fig. 2(d), the (0.5 0.5 1.5) reflection increases monotonically with electron doping, leading to a substantially distorted lattice. Because octahedral tilting originates from the deviation of the out-of-plane Ir-O-Ir bond angle from 180° (inset of Fig. 2(d)), these results demonstrate that the out-of-plane bond angle is strongly reduced due to La substitution.

We note that the residual interlayer coupling is essential for stabilizing the quasi-2D AFM long-range order in the undoped SLs[25,42] as well as in other low-dimensional magnets[43,44]. For the SL family, this coupling is mediated by electron hopping across the $SrTiO_3$ spacer and can therefore be weakened by octahedral tilting[23,24,45]. The enhanced octahedral tilting in the electron-doped samples may therefore enable the rapid reduction of $T_N$. In a correlated Mott system with intermediate correlation strength, however, a reduction of $T_N$ is generally accompanied by a decrease in the charge gap[46], consistent with previous observations that epitaxial strain-enabled structural modulation simultaneously tunes $T_N$ and charge gap[29]. Therefore, the enhanced average octahedral tilting provides a plausible contribution to the rapid suppression of both $T_N$ and $\Delta$ in the electron-doped regime, but does not account for their pronounced decoupling on the hole-doped side, where the average lattice distortions remain comparatively unchanged.

Note that the SXRD measurement cannot completely exclude additional local structural effects associated with the distinct chemical substitution routes, such as nanoscale and disordered distortions surrounding the dopants. Investigating this possibility will require advanced structural and magnetic measurements with atomic-scale spatial resolution[47,48]. Nevertheless, the similar electron-hole asymmetric behavior found in *A*-site substituted cuprates[31-34], strongly argues against the observed asymmetry being solely a consequence of the different substitution sites or average structural distortions.

**Emergent hidden order in the hole-doped regime**

To shed further light on the decoupling between $T_N$ and $\Delta$, we directly characterize the magnetic structure by combining magnetization and REXS measurements. Owing to its element specificity and ultrahigh sensitivity, REXS is particularly suitable for this study.

Figure 3(a) shows the energy dependence of the (0.5 0.5 1) magnetic reflection near the Ir-$L_3$ edge. The pronounced resonance at this reflection demonstrates that all SLs host the same *C*-type AFM structure, with the magnetic order arising from the Ir ions (see Supplementary Notes 5-7). Note that the magnetic reflection already approaches the detection limit at -4% (see Supplementary Notes 6-7), the magnetic reflection of the -8% sample is therefore expected to be well below the experimental resolution of our REXS setup. Furthermore, in single-layered iridates, such as $Sr_2IrO_4$ and $Ba_2IrO_4$, it was unveiled that the in-plane magnetic anisotropy is remarkably robust against tetragonal distortion, and is further stabilized by interlayer interactions[49-51], which are also substantial in our single-layered iridates[24]. Finite octahedral tilting further favors the in-plane magnetic anisotropy[52], and suppresses the *c*-axis anisotropy[45] frequently observed in bilayer iridates as a consequence of dimensionality effect[53-55]. We therefore expect the doped SLs to preserve the same in-plane magnetic anisotropy as the pristine sample. In addition, unlike $Sr_2IrO_4$, where carrier doping can modify the interlayer interaction and even induce a reorientation of the net magnetization of individual $IrO_2$ layers[56,57], the SLs have a substantially simpler magnetic structure which retains a ferromagnetic interlayer interaction throughout the entire doping series.

Therefore, the spin canting induced magnetization would be proportional to the magnitude of Neel order $\boldsymbol{L}$ and the spin canting angle $\phi$ in the whole SL series (inset of Fig.

3(b))[13,15,50,58]. At zero magnetic field, ϕ equals to the octahedral rotation angle[26,42,59], which can be determined from the square root of the intensity $\boldsymbol{I_C}$ of the relevant superstructure peaks, such as the (0.5 1.5 1.5) reflection in Fig. 2(a)[40,41]. The averaged $\boldsymbol{L}$ of each SL was then extracted by $\bar{\boldsymbol{L}} = \boldsymbol{A}\times\boldsymbol{M}/\sqrt{\boldsymbol{I_C}}$, where $\boldsymbol{M}$ is the net magnetization in Fig. 1(b). The constant A was taken as 1 for simplicity. As summarized in Fig. 3(b), the base-temperature $\bar{\boldsymbol{L}}$ decreases continuously with electron doping, while it first increases and then decreases continuously with hole doping.

In REXS measurements, the square root of the magnetic peak intensity $\boldsymbol{I_M}$ is also governed by $\bar{\boldsymbol{L}}$[15,26,51,60]. Specifically, $\sqrt{\boldsymbol{I_M}}$ was proved to strictly follow $\bar{\boldsymbol{L}}$ in pristine SLs[26,42]. We then scaled $\sqrt{\boldsymbol{I_M}}$ to $\bar{\boldsymbol{L}}$ for the pristine SL and included $\sqrt{\boldsymbol{I_M}}$ of other SLs using the same scaling factor in Fig. 3(b). In the electron-doped SLs, the scaled $\sqrt{\boldsymbol{I_M}}$ is slightly weaker than $\bar{\boldsymbol{L}}$. Notably, $\sqrt{\boldsymbol{I_M}}$ surprisingly experiences an upturn before eventually gets suppressed due to hole doping. The intriguing variation of $\sqrt{\boldsymbol{I_M}}$ in the hole-doped SLs is further confirmed at different beamline stations and/or with different analyzing strategies (see Supplementary Notes 8 and 9). In particular, the scaled $\sqrt{\boldsymbol{I_M}}$ is about twice of $\bar{\boldsymbol{L}}$ at the hole doping ratio ~9%. This increase is sufficiently large to exclude the same dominant role of the AFM order in $\sqrt{\boldsymbol{I_M}}$ and $\bar{\boldsymbol{L}}$, and suggests that the percolation type transition as observed in Rh-doped $Sr_2IrO_4$ is unlikely to occur in the Rh-doped SLs[56].

In addition to spin reorientation, which is unlikely based on the considerations above, changes in the resonant scattering factor of dipole moments could also modify the REXS intensity. However, the Hall measurements indicate only a modest change in carrier density (see Supplementary Note 10), while the Ir $L_3$ edge XAS spectra show essentially unchanged Ir valence within experimental resolution (see Supplementary Note 5). A substantial change in the resonant scattering factor is therefore unlikely to account for the pronounced enhancement of the REXS intensity. These observations rather suggest that the enhanced REXS intensity is more naturally associated with the presence of another order that appears on top of the conventional AFM order in the hole-doped SL. This order would be staggered on neighboring Ir sites akin to the AFM order, but is fully compensated so that it does not contribute to the net magnetization.

To verify this scenario, we investigated temperature evolutions of $\sqrt{\boldsymbol{I_M}}$ and $\bar{\boldsymbol{L}}$. The onset temperature of $\sqrt{\boldsymbol{I_M}}$ is labeled as $T_\Omega$, which equals to $T_N$ in the pristine SL (Fig. 3(c))[26]. Interestingly, in the electron-doped SLs, $T_\Omega$ remains close to $T_N$, with their difference falling within the experimental uncertainty. In contrast, when introducing additional holes, $T_\Omega$ is systematically higher than $T_N$ throughout the investigated doping range, and their separation increases monotonically with the hole-doping ratio. This deviation reaches to over 20 K in the 14% doped SL (Fig. 3(c)). As summarized in Fig. 3(d), $T_\Omega$ follows $T_N$ in the electron-doped regime within the experimental error but decouples in the hole-doped regime. Note that, for a given order parameter, different experimental probes sensitive to the same order should identify the same transition temperature. The observation that $T_\Omega > T_N$ in the hole-doped regime further suggests that another order may develop prior to the onset of the conventional AFM order.

**Effects of the hidden order on electron transport**

We noted that both the hidden order and the decoupling between $T_N$ and $\Delta$ have the same electron-hole asymmetry, *i.e.*, both phenomena can only be observed in the hole-doped regime. This similarity indicates that the hidden order may play an important role in the decoupling between $T_N$ and $\Delta$. We then further investigated the spatial modulation of magnetic orders in the doped SLs. The in-plane coherence length $\xi_{ip}$ was extracted from *H*-scan of the (0.5 0.5 1) reflection following the relation $\xi_{ip} = 1/(\frac{\text{FWHM}}{2} \cdot 2\pi)$, where FWHM is the full width at the half maximum of the *H*-scan. As shown in Fig. 4(a), $\xi_{ip}$ decreases substantially and monotonically with electron doping, unveiling a strongly weakened magnetic correlation. On the other hand, $\xi_{ip}$ only slightly reduces upon doping with 3% of hole, and remains almost unchanged with further increasing the hole-doping ratio. Strikingly, $\xi_{ip}$ in the -3% electron-doped SL is only about one third of that in the 12% hole-doped SL, although both SLs having similar $T_N$ values (Fig. 3(d)). To determine the crystalline coherence length, we extracted the FWHM of the (0.5 1.5 1.5) peak, which is strong and well-resolved throughout the entire series. Interestingly, the crystalline coherence length shows only a weak dependence on doping, indicating a comparable crystalline quality across the SL series (inset of Fig. 4(a)). The contrasting doping dependences of the magnetic and crystalline coherence lengths suggest that a doping-induced structural transition is unlikely to account for the preserved magnetic coherence in the hole-doped regime.

In addition, the out-of-plane coherence length $\xi_{oop}$ of the pristine SL is about 20 unit cells (u.c.). This value is limited by the ultrathin nature of the SLs and is quantitatively consistent with previous reports[23,25]. More importantly, similar to $\xi_{ip}$, $\xi_{oop}$ is also substantially reduced in the electron-doped regime, but is almost unchanged in the hole-doped regime (Fig. 4(b)). Therefore, the REXS measurements unveiled that magnetic coherence is significantly disturbed due to electron doping, while it is largely preserved upon hole doping in the iridates. The observation of well-defined magnetic reflections at $T_N$ further supports the presence of an additional order that extends beyond the conventional AFM transition and may contribute to the preserved magnetic coherence in the hole-doped regime (see Supplementary Note 11). Notably, the preserved magnetic coherence (Fig. 4(a)), the enhanced $\sqrt{I_M}$ (Fig. 3(b)), and the distinct $T_\Omega > T_N$ (Fig. 3(c)) emerge exclusively in the hole-doped regime. In particular, $\sqrt{I_M}$, $T_\Omega$ and the magnetic coherence length are extracted from the same magnetic Bragg reflection and thus probe the same magnetic ordering wavevector. This suggests that they may be different manifestations of a common hidden order. Nevertheless, the detailed doping dependences of the REXS intensity and $T_{\Omega}$ do not exactly coincide, and the possibility of intertwined or multiple ordering phenomena cannot be completely excluded.

Having established the anomalous modulation of magnetic coherence upon hole doping, we next examine its possible relationship with electronic transport. In the pristine SL, the AFM Mott insulating ground state is well established in layered iridates[9,10,14,61,62]. Carrier doping weakens this state, allowing carriers to travel over finite distances. Because the insulating state is closely associated with the AFM order, the strongly reduced magnetic coherence in the electron-doped regime is consistent with spatially nonuniform hopping barriers (left panel of Fig. 4(c)). In this context, the global carrier mobility may therefore remain relatively low, leading to similar doping dependences of $T_N$ and $\Delta$. In contrast, the anomalously preserved magnetic coherence in the hole-doped regime is consistent with hole transport on a more coherent and uniform background (right panel of Fig. 4(c)). The concomitant increase in carrier density and mobility can then account for the faster suppression of $\Delta$ than $T_N$. With increasing the Rh substitution ratio, the possible presence of chemical clustering or local chemical inhomogeneity may counteract the mobility enhancement associated with the hidden order, potentially accounting for the saturation of $\Delta$ at the highest hole-doping level. Within this picture, the distinct evolution of magnetic coherence and electronic transport, together with the emergence of the hidden order exclusively in the hole-doped regime, provides a possible framework for understanding the pronounced electron-hole asymmetry in the coupling between $T_N$ and $\Delta$.

We note that a glass-like phase separation lacking a long-range order has been visualized in electron-doped $Sr_2IrO_4$[21], and the insulating state breaks into short-ranged fragments near the insulator to metal transition[18]. Experimentally, the slightly smaller $\sqrt{\boldsymbol{I_M}}$ than $\bar{\boldsymbol{L}}$ in the left side of Fig. 3(b) is consistent with the rapid degradation of magnetic coherence upon electron doping, suggesting that short-range magnetic domains or phase-separated regions may also develop in the electron-doped SLs. By contrast, holes in Rh-doped $Sr_2IrO_4$ are more itinerant than electrons in La-doped $Sr_2IrO_4$[63]. These observations are consistent with the distinct evolution of magnetic coherence and charge transport observed here and suggest that the different roles of electrons and holes may be a general feature of doped layered iridates. Importantly, however, the present measurements do not establish the direction of causality between hidden order and $T_N$-$\Delta$ decoupling. The hidden order could contribute to the anomalous $T_N$-$\Delta$ relationship, while an alternative scenario in which the underlying electronic and magnetic correlations promote the hidden order cannot be excluded.

Previously, various hidden orders have been reported or proposed in doped or pristine $Sr_2IrO_4$[2,16,17,19,64-66]. In contrast to the hidden orders reported in both pristine and carrier-doped layered iridates[17,21,65,66], the hidden order observed here emerges exclusively in the hole-doped regime. One possible microscopic origin may be related to the particular $t_{2g}$ character of the $J_{eff} = 1/2$ manifold that underlies various magnetic states and excitations[67-69]. Within the $J_{eff} = 1/2$ doublet of pristine iridates, magnetic operators reduce to dipolar moments, suppressing the formation of higher-rank multipoles, and only when intersite hopping processes are considered can multipolar order arise[19]. Upon electron doping, the occupied $t_{2g}$ shells further limit the possibility of higher-rank magnetic operators[70,71]. In contrast, hole doping may enlarge the accessible local Hilbert space[70] and allow additional higher-rank operators to become active, potentially favoring multipolar correlations. This scenario could be further examined by high-brilliance, high-resolution REXS measurements, which may resolve characteristic energy-dependent signatures of different multipolar channels[72,73].

On the other hand, the presence of a qualitatively similar electron-hole asymmetric relationship between $T_N$ and the insulating strength in cuprates, despite their simpler $S = 1/2$ local Hilbert space, suggests that this asymmetry may reflect a more general electronic feature of doped Hubbard systems rather than being associated with the complexity of the $t_{2g}$ manifold. This observation motivates further investigation of electron-doped cuprates to determine whether analogous hidden order exists and, if so, whether it shares the same microscopic origin as iridates.

In summary, we observed a pronounced electron-hole asymmetry in the coupling between the insulating strength and $T_N$ in iridate SLs. While enhanced octahedral tilting under electron doping can account for the rapid degradation of the AFM insulating state, it is unlikely to be the dominant player governing the distinct relationship between insulating strength and $T_N$ in the electron- and hole-doped regimes. Moreover, in the electron-doped SLs, magnetic coherence is rapidly suppressed, consistent with the rapid reduction of the AFM order. In contrast, magnetic coherence is anomalously robust in the hole-doped regime, accompanied by an enhanced REXS response and a distinct $T_\Omega > T_N$, indicating the presence of a hidden order that is associated with the AFM state but does not contribute to the net magnetization. The magnetic nature of this hidden order may be responsible for the absence of superconductivity in Rh-doped layered iridates.

## Methods

### Superlattice growth

$(La_xSr_{1-x}IrO_3)1/(SrTiO_3)1$ and $(SrIr_{1-y}Rh_yO_3)1/(SrTiO_3)1$ SLs were deposited on (001)-oriented $SrTiO_3$ (STO) substrates by the pulsed laser deposition (PLD) technique. STO substrates were treated with HF solution and annealed at 1050°C prior to growth. A reflection high-energy electron diffraction (RHEED) system was utilized to monitor the growth. During deposition, $O_2$ was introduced at a constant partial pressure of 75 mTorr, and the growth temperature was maintained at 700 °C for all the SLs. PLD was carried out using a KrF excimer UV laser (λ = 248 nm), with the energy density of 1.2 $J/cm^2$ and repetition rate of 3 Hz. The doping ratio was controlled by counting the number of pulses required for the nominal composition, as detailed in a previous work[74]. For the lightly doped SLs, the required pulse number was close to the experimental limit. After deposition, the films were cooled to room temperature at a rate of 10 °C/min under an oxygen partial pressure of about 350 Torr.

ASSOCIATED CONTENT

**Data availability.** The data that support the findings of this study are shown in the main text figures and the supplementary figures.

**Supporting Information.** The Supporting Information is available free of charge, and includes Supplementary Notes 1–11, Figures S1–S10, and Table S1.

AUTHOR INFORMATION

**Author contributions**

Conceptualization: LH; Methodology: ZQ, Han Zhang, LH; Investigation: ZQ, Han Zhang, DL, Haiyang Zhang, EM, HW, JJ, JC, YL, LH, XX, NH; Resources: SYP, MLT, HD; Funding acquisition: LH; Project administration: LH; Supervision: LH; Writing – original draft: Han Zhang, LH; Writing – review & editing: ZQ, Han Zhang, LH. All authors have read and approved the manuscript.

ACKNOWLEDGEMENTS

We appreciate experimental assistances from Xin Zhang and Hironori Nakao. We also thank Zhentao Wang, Liang Si and Tae Heon Kim for helpful communications. A portion of this work was performed on the Steady High Magnetic Field Facilities, High Magnetic Field Laboratory, Chinese Academy of Sciences, and supported by the High Magnetic Field Laboratory of Anhui Province. A portion of this work was performed under the approval of the Photon Factory Program Advisory Committee (Proposal No. 2023G509).

**Funding**

This work was supported by the National Key Basic Research Program of China (2024YFA1611100), the National Natural Science Foundation of China (12104460 and 12474125), the international partnership program of the Chinese Academy of Sciences (145GJHZ2022044MI), the Anhui Provincial Major S&T Project (s202305a12020005), and the Basic Research Program of the Chinese Academy of Sciences Based on Major Scientific Infrastructures (JZHKYPT-2021-08). Han Zhang acknowledges the support from National Natural Science Foundation of China (grant No. 12574033, 12204068) and the China Postdoctoral Science Foundation (grant No. 2025T180149, 2024M753261). The funders had no role in the study design, data collection and analysis, interpretation of the results, or preparation of the manuscript.

## Competing interests

The authors declare no competing financial or non-financial interests.

**Figures**

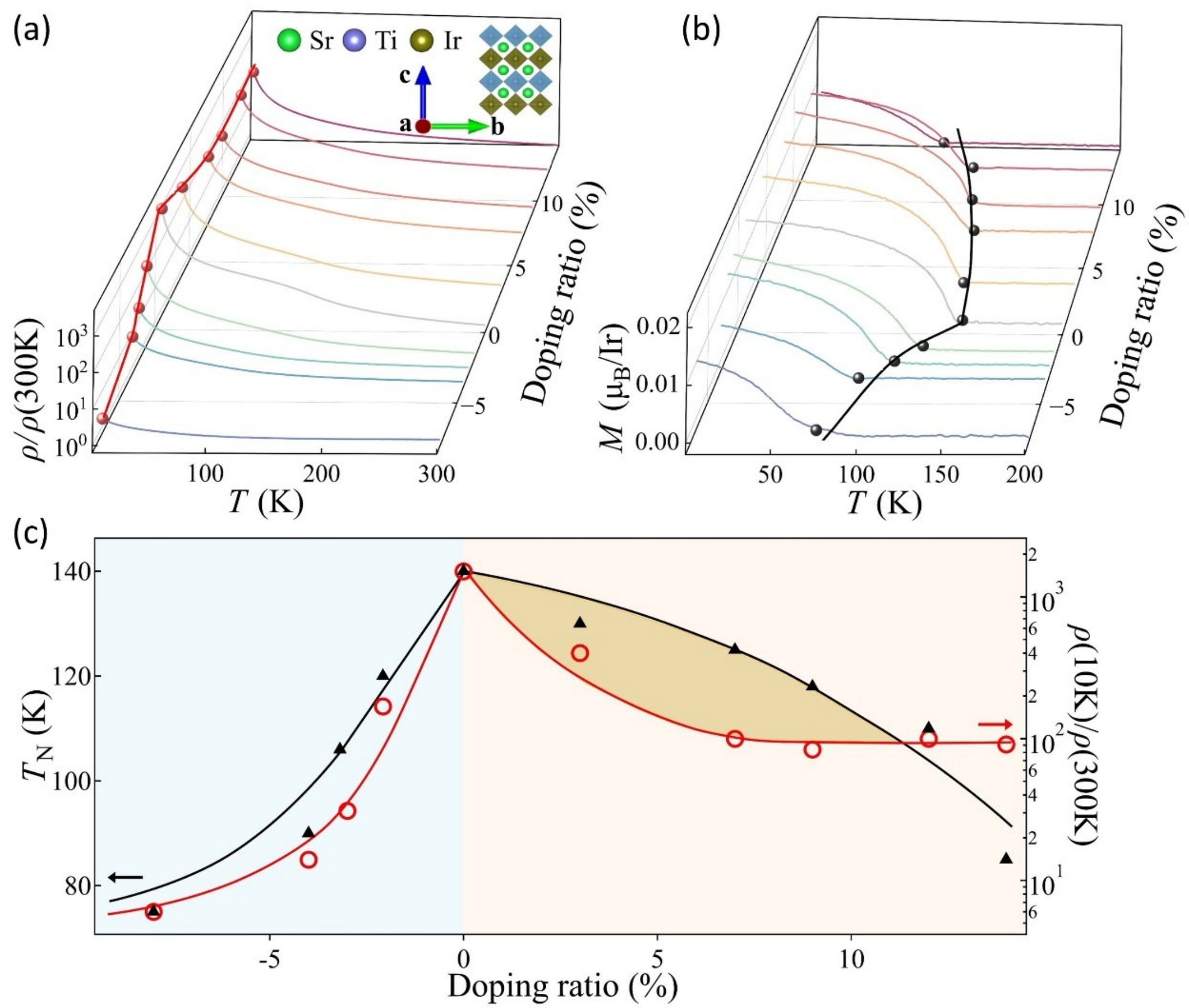

**Fig. 1. Electron-hole asymmetry in iridates. (a)** Temperature-dependent normalized resistivity of the carrier-doped SLs. The inset displays the crystalline structure of the pristine SL. **(b)** Temperature-dependent magnetization of the SLs. The magnetization was measured at zero field after cooling the samples in a 0.5 T in-plane magnetic field. **(c)** $T_N$ (left) and insulating strength (right) as functions of doping ratio. The insulating strength is defined by $\rho(10K)/\rho(300K)$.

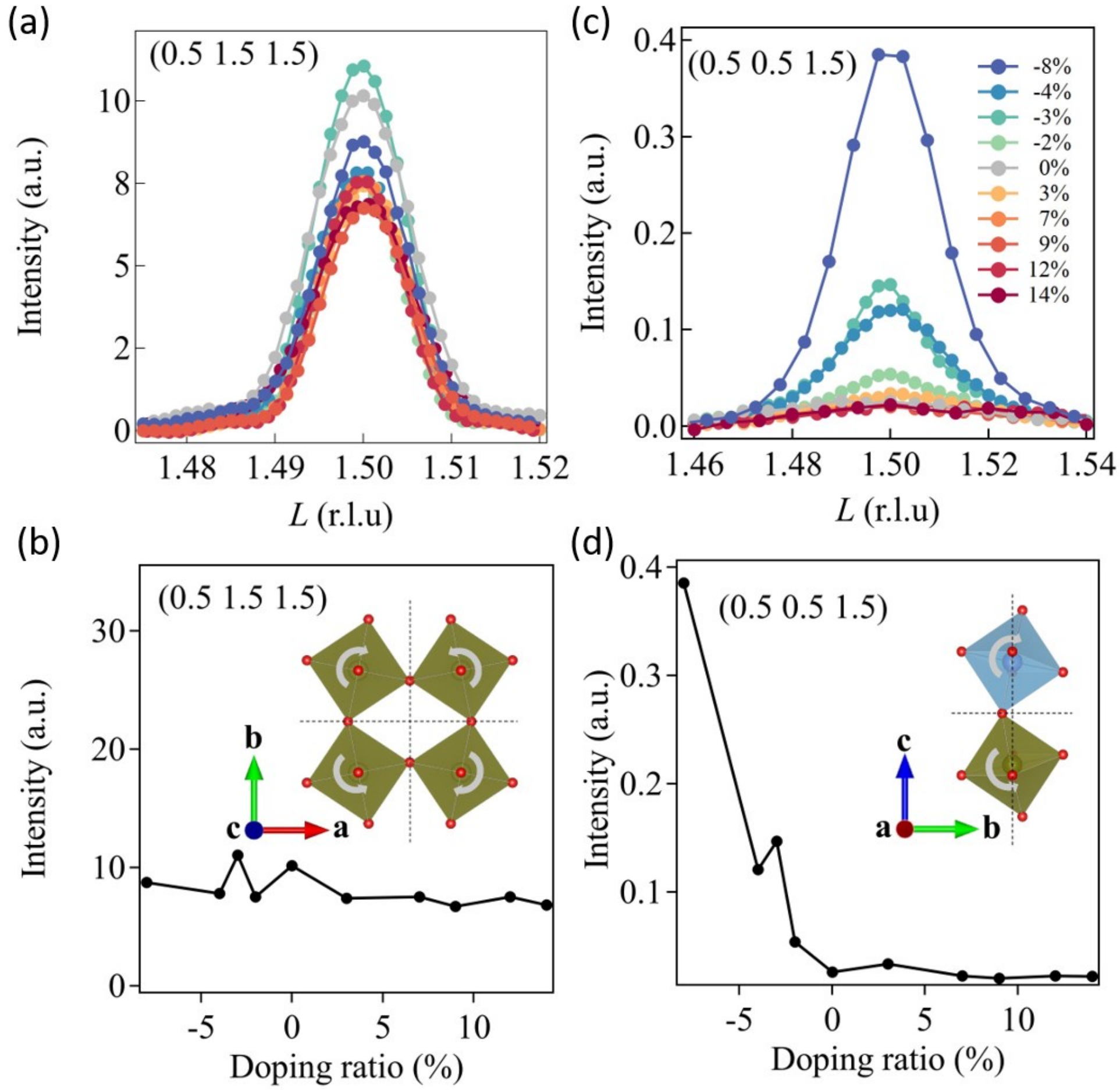

**Fig. 2. Characterization of the octahedral rotation pattern. (a)** SXRD around the (0.5 1.5 1.5) peak for all the SLs. **(b)** Dependence of the (0.5 1.5 1.5) peak intensity on the doping ratio. Inset shows a schematic diagram of octahedral rotation. **(c)** SXRD around the (0.5 0.5 1.5) reflection. The reciprocal space was defined in the pseudo-cubic cell of each SL. **(d)** Dependence of the (0.5 0.5 1.5) peak intensity on the doping ratio. Inset shows a schematic diagram of octahedral tilting.

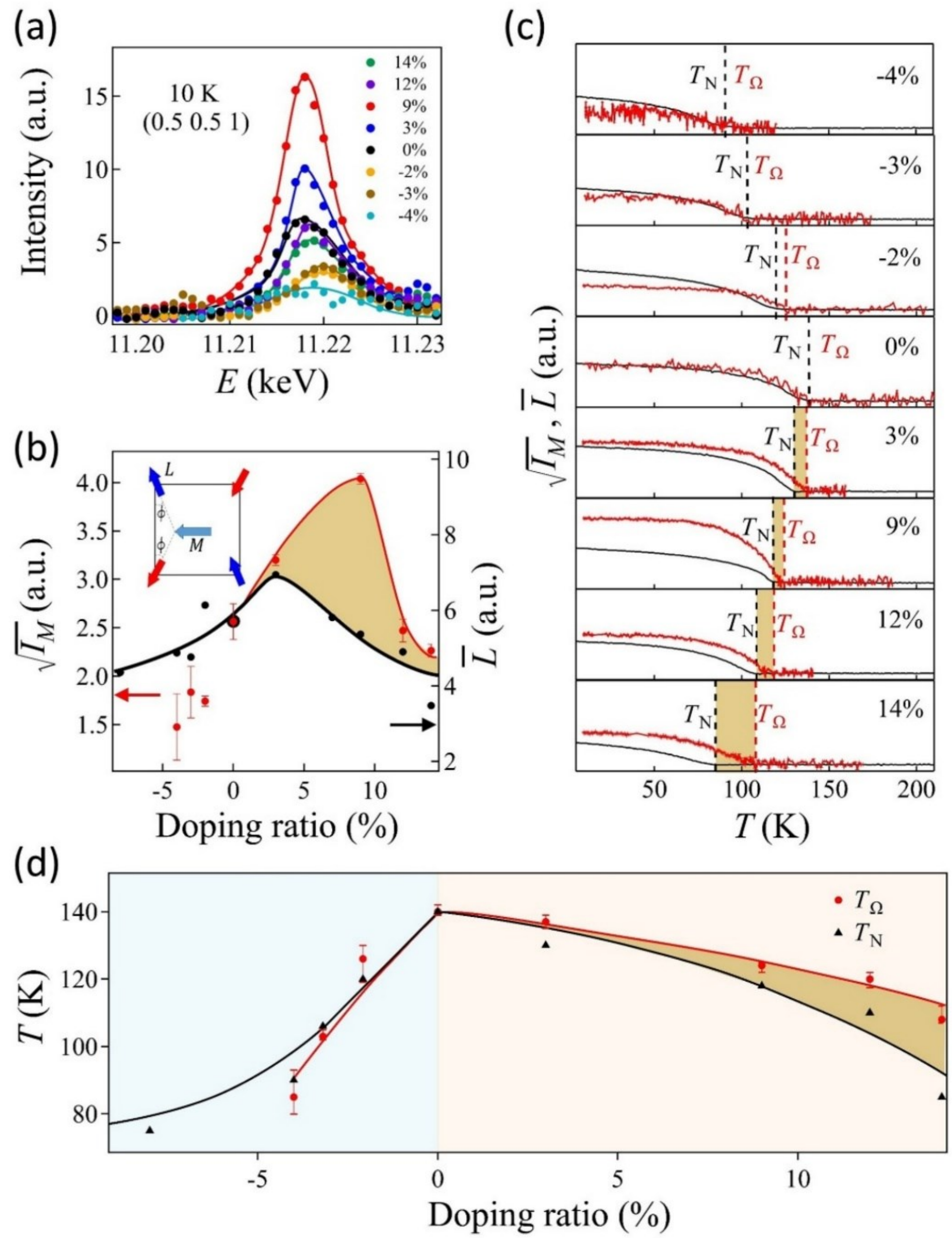

**Fig. 3. Hidden order in hole-doped iridates. (a)** Energy scan at the (0.5 0.5 1) magnetic reflection around the Ir $L_3$ edge at 10 K. **(b)** $\bar{\boldsymbol{L}}$ (black circles) and $\sqrt{\boldsymbol{I_M}}$ (red circles) as functions of the doping ratio. Solid curves are guides for the eye. The inset displays a schematic illustration of the spin-canted AFM structure. **(c)** Temperature-dependent $T_N$ and $T_\Omega$ of various SLs. The discrepancy between $T_N$ and $T_\Omega$ is highlighted by a light-brown region. **(d)** Doping-ratio-dependent $T_N$ and $T_\Omega$ of SLs.

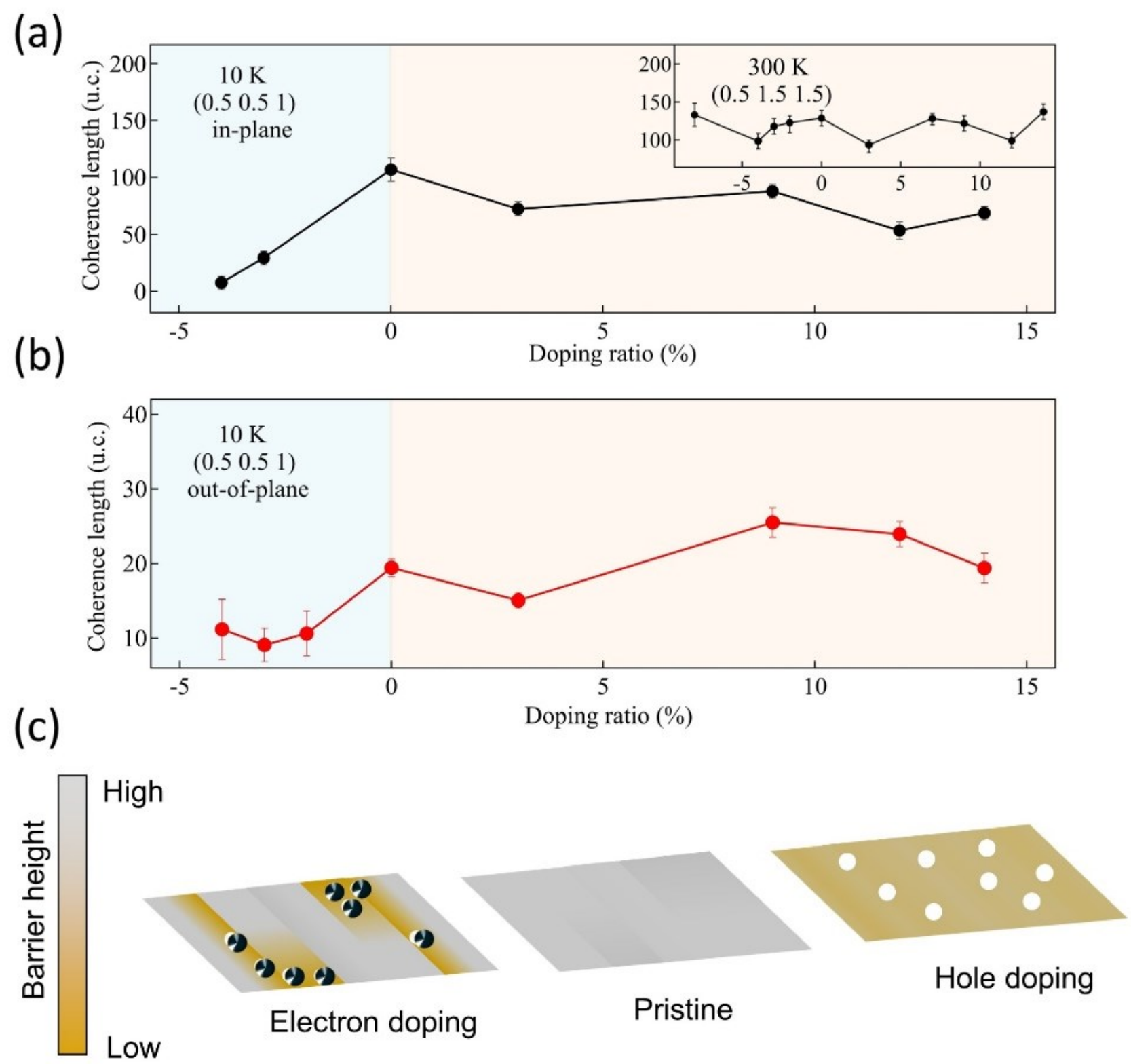

**Fig. 4. Unconventional modulation of magnetic coherence upon carrier doping.** Doping-dependent $\xi_{ip}$ **(a)** and $\xi_{oop}$ **(b)** at 10 K. The inset displays the relation between in-plane crystalline coherence length and doping ratio. **(c)** Schematic diagrams of the electronic transport in electron-doped (left), pristine (middle), and hole-doped (right) regimes. The low-barrier and high-barrier regions are sketched as bronze and grey regions, respectively.